\documentclass[hidelinks,12pt]{article}
\usepackage{bm}
\usepackage{amsmath, amsthm, amssymb}
\usepackage{setspace}
\usepackage{multirow}
\usepackage{graphicx}
\usepackage[square,sort,comma,numbers]{natbib}
\usepackage{hyperref}
\usepackage{mathrsfs}
\usepackage{mdwlist}
\usepackage{rotating}
\usepackage{arydshln}
\usepackage{pdflscape}

\hypersetup{
    colorlinks=false,
    pdfborder={0 0 0},
}

\newcommand{\beq}{\begin{equation}}
\newcommand{\eeq}{\end{equation}}

\newcommand{\captionfonts}{\footnotesize}
\makeatletter
\long\def\@makecaption#1#2{
  \vskip\abovecaptionskip
  \sbox\@tempboxa{{\captionfonts #1: #2}}
  \ifdim \wd\@tempboxa >\hsize
    {\captionfonts #1: #2\par}
  \else
    \hbox to\hsize{\hfil\box\@tempboxa\hfil}
  \fi
  \vskip\belowcaptionskip}
\makeatother

\begin{document}

\title{A comparative analysis of the UK
and Italian small businesses using
Generalised Extreme Value models}
\author{Galina Andreeva$^{1}$, Raffaella Calabrese$^2$ and Silvia Angela Osmetti$^3$\\
\small \small $^1$Business School, University of Edinburgh\\ \small
29 Buccleuch Place, Edinburgh EH8 9J, UK\\ \small \texttt{Galina.Andreeva@ed.ac.uk} \\
\small $^2$Essex Business School, University of Essex\\ \small
Wivenhoe Park, Colchester CO4 3SQ, UK\\ \small \texttt{rcalab@essex.ac.uk} \\
\small $^3$Department of Statistical sciences, Universit\`{a} Cattolica del Sacro Cuore di Milano \\ \small Largo Gemelli 1, 20123 Milano, Italy \\ \small  \texttt{silvia.osmetti@unicatt.it}
}

\date{ }
\maketitle
\noindent
Corresponding author: Dr Galina Andreeva\\
Tel:+44(0)65123293\\
Fax:+44 (0)131 651 3197\\
E-mail: Galina.Andreeva@ed.ac.uk

\pagebreak

\begin{abstract}

This paper presents a cross-country comparison of significant predictors of small business failure between Italy and the UK. Financial measures of profitability, leverage, coverage, liquidity, scale and non-financial information are explored, some commonalities and differences are  highlighted. Several models are considered, starting with the logistic regression which is a standard approach in credit risk modelling. Some important improvements are investigated.  Generalised Extreme Value (GEV) regression is applied to correct for the symmetric link function of the logistic regression. The assumption of non-linearity is relaxed through application of BGEVA, non-parametric additive model based on the GEV link function. Two methods of handling missing values are compared: multiple imputation and Weights of Evidence (WoE) transformation. The results suggest that the best predictive performance is obtained by BGEVA, thus implying the necessity of taking into account the relative volume of defaults and non-linear patterns when modelling SME performance. WoE for the majority of models considered show better prediction as compared to multiple imputation, suggesting that missing values could be informative and should not be assumed to be missing at random.

 \end{abstract}
\vspace{0.4cm}
\noindent \textbf{Keywords:} Decision support systems, Risk analysis, Credit Scoring, Small and Medium Sized Enterprises,
Default prediction.

\section{Introduction}

Small and Medium Enterprises (SMEs) play a central role in the European Union (EU) economy, as recognised by the
Small Business Act of the European Commission in 2008 (http://ec.europa.eu/enterprise/entrepreneurship/docs/sba/SBA$\_$IA). In 2011 SMEs represented  99\% of enterprises in Europe, employing more than two  thirds of the workforce and contributing  58\% of total EU added value. The importance of SMEs varies across the EU. In some countries, e.g. Italy, Spain and Portugal, SMEs have larger shares in employment and added value and higher presence than the EU average. On the contrary, these figures are lower than the EU average in other countries, e.g. the UK, Germany and France.

In this work we compare Italy and the UK since the economies of these countries are different, and it is of interest to explore the differences in predictors of SMEs failures, especially in the aftermath of the "credit crunch". The literature on SME default prediction is limited, in particular in cross-country comparisons, and the main objective of this paper is to fill in this gap.
This paper contributes to the existing cross-country research by an initial exploratory investigation of risk predictors using accounting and some non-financial information that are available from public sources.

Several models are considered, starting with the logistic regression which is a standard modelling approach in credit risk research \citep{Thomas:2002}. Yet despite its popularity,  logistic regression has some limitations that can be of importance in credit risk situations. In this paper we concentrate on symmetric link function and the assumption of linearity between the response and predictors. In real applications the proportion of defaults is small, therefore, making the symmetric link function questionable. Equally the assumption of linearity is not always supported by patterns in the real data. An additional contribution of this paper consists in extending the application of Generalised Extreme Value (GEV) regression that has been proposed for low  default portfolios by \cite{Calabrese:2013} to two countries. Furthermore, the problem of non-linearity is explored  through the application of non-parametric additive model (BGEVA).

The public sources often have incomplete data and this problem is particularly relevant for SMEs. Another objective and contribution of this paper consists in exploration of two approaches to handling the missing values: multiple imputation and Weights of Evidence transformation which is credit industry's preferred approach.

The rest of the paper is structured as follows: Section 2 provides some background information on the importance of SMEs to the economy and some differences across the two countries. It also summarises previous research on SMEs failure prediction. Section 3 explains the methodology, and Section 4 presents the empirical results, including data description, comparison of predictive accuracy and comparison of statistically significant risk predictors. The final section concludes.
\section{Background and literature review}

There are some notable differences in characteristics of SMEs in the UK and Italy. In Italy, SMEs form $99.9\%$ of the firms. In 2011 they employed around $81\%$ of the workforce and contributed $68.3\%$ of the Italian added value \citep{European:2012}. In terms of the number of SMEs, Italy has the largest SME sector in the EU. With 3.813 million SMEs Italy has almost twice as many as UK (1.649 million). However, the vast majority of Italian SMEs are micro-firms with less than 10 employees. In fact, Italy's share of micro-firms, at 94.6\%, exceeds the EU-average (92.2\%). Hence, the micro-firms' contribution to employment (46.6\%  against the EU-average of 29.6\%) and added value (29.4\% against the EU-average of 21.2\%) is high.

On the contrary, the UK economy is characterised by larger companies. In 2011 more than half of the UK added value was produced by large companies that employed less than half (45.7\%) of the workforce and constituted only 0.4\% of the UK companies. The percentage of micro-firms in the UK (89.5\%) is lower than the EU-average (92.2\%), and those employ only 20.3\% of the workforce and create only 18.5\% of the UK added value \citep{EuropeanUK:2012}.

Financial crisis has substantially affected SMEs sectors in both countries and recovery has been weaker than in the EU on the whole. The Italian SME sector has reversed to the levels of 2005 (i.e. before the crisis) in terms of the number of firms, employment and value-added creation. In the UK, SMEs have been hit mostly in terms of employment and value-added creation, but the numbers of SMEs are higher than in 2005 and stable. In both countries larger firms suffered less as compared to the smaller ones.

Despite an important role that SMEs play in any economy, academic research into SMEs failure prediction is not very extensive. There are some (albeit not numerous) papers investigating success factors or default risk of SMEs in a specific country, e.g. \cite{Altman:2007} for the US, \cite{Fantazzini:2009} for Germany, \cite{Sohn:2013} for South Korea, \cite{Martens:2011} for Flanders - to give some examples, yet literature on international comparisons of failure prediction is exceptionally limited.

The survey by \cite{Altman:1997} summarised previous research on the performance of companies (not only SMEs) in 22 countries that included both developed and developing economies. Most studies surveyed found measures of profitability, leverage, liquidity, cash flow management, growth, efficiency to be important for bankruptcy prediction, although specific measures used would vary from country to country. More recent study by \cite{Lussier:2010} compared performance of SMEs in the USA, Croatia and Chile. Among the variables that were found important for business performance were characteristics of managers (education, experience) and the quality of business functions (record keeping, financial control, planning, staffing).

The most comprehensive study of European SMEs to date is by \cite{Michala:2013} where a simple hazard model \citep{ Shumway:2001} has been applied to small businesses from eight European countries, namely Czech Republic, France, Germany, Italy, Poland, Portugal, Spain and the United Kingdom for the period of 2000-2009. The paper has confirmed the significance of indicators of profitability, coverage, leverage and cash flow for bankruptcy prediction in cross-country setting. In addition, some non-financial company characteristics have been investigated and the effect of macroeconomic variables. \cite{Pederzoli:2013} modelled credit risk of EU innovative SMEs, but the authors did not make cross-country comparisons.

There were some comparisons between two countries. \cite{Ihua:2009} compared the key factors influencing SMEs failure between the UK and Nigeria, and found that economic conditions and infrastructure were more significant in Nigeria, whilst in the UK the key factors were due to internal company characteristics, including management efficiency.

\cite{Dietsch:2004} analysed default probabilities and asset correlations for French and German SMEs. Yet the focus of their analysis was more on comparison of correlations of SMEs as opposed to large corporations, the paper did not look at financial ratios or other predictors of default.

As for SME research in the UK, \cite{Lin:2012} compared different definitions of financial distress on sample from 2001 to 2004 and concluded that although each definition changed the model composition substantially, the most useful variables in distinguishing between distressed and healthy companies, were profit related measures, growth and efficiency ratios.
\cite{Altman:2010} developed a default prediction model using financial indicators of leverage, profitability, working capital and non-financial information (e.g. age, default events in the past) using the data from 2000 to 2007. They found the non-financial variables provided a notable improvement in predictive performance. \cite{Orton:2011} explored the behaviour of the UK SMEs from 2007 to 2010 - through the "credit crunch"�.  They demonstrated that there was a significant degree of stability and accuracy of credit risk models, despite increases in the numbers of SMEs defaults. Similar to \cite{Altman:2010} they found company demographics, derogatory events and information about directors to be of significant value.

Regarding the modelling approaches, the overwhelming majority of studies reviewed above used logistic regression. Other models included proportional odds or simple hazard model \citep{Michala:2013, Figini:2009}, Bayesian and classic panel models \citep{Figini:2009}, random survival forests \citep{Fantazzini:2009}, Support Vector Machines \citep{Martens:2011}.

In Italy \cite{Vallini:2009} attempted to model SME defaults on a sample of small firms from 2001- 2005 using profitability, liquidity and leverage ratios. Multiple discriminant analysis was compared to logistic regression, and the latter was found to produce better predictions. Later study by \cite{Ciampi:2013} applied neural networks to the same dataset and reported their superior performance as compared to algorithms used in the earlier work. Both studies noted that credit scoring models could be built on accounting information, yet predicting default for SMEs was much more difficult as compared to large enterprises, with predictive accuracy decreasing in smaller firms segments.

\cite{Calabrese:2013} and \cite{Marra:2013} applied GEV and BGEVA models to the sample of  Italian SMEs from 2006 to 2011 and found superior performance of both models as compared to logistic regression. Variables found significant in predicting default were again measures of profitability, leverage and liquidity.

The current paper extends the existing literature by looking at two countries in comparison  (Italy and the UK), by exploring SMEs failure in a more recent time period and by using more comprehensive list of financial measures.

\section{Methodology}

When constructing a credit scoring model, three common problems are often mentioned: first, the symmetric link function of regression models commonly used in credit scoring, second, non-linear relationship between the response and predictors, and third, missing values in predictor variables.

 Logistic regression is the most commonly used model for credit scoring applications \citep[e.g.,][]{Altman:2007,Becchetti:2002,Lin:2012,Zavgren:1998}. Since the number of defaults in a sample is usually very small \citep[e.g.,][]{Kiefer:2010,Lin:2012}, the use of the logit link function may not be appropriate because of its symmetry around $0.5$. This is not ideal as the characteristics of defaults are more informative than those of non-defaults and as a consequence the probability of default could be underestimated \citep{Calabrese:2013}. Moreover, the bias of maximum likelihood parameter estimators for logistic regression is amplified when the number of defaults is low \citep{King:2001}. This suggests using an asymmetric link function as in \cite{Wang2010}.

In order to choose the link function, we consider that defaulters' features are represented by the tail of the response curve for values close to one. Furthermore, the Generalised Extreme Value (GEV) distribution is used in literature \citep{Kotz:2000,Falk:2010} to model the tail of a distribution. Therefore, to focus the attention on defaulters' characteristics, \cite{Calabrese:2013} propose the quantile function of a GEV random variable as a new link function
\begin{equation}
\label{linkgev}
\frac{\left[-\ln (PD_i)\right]^{-\tau}-1}{\tau}=\eta_i=\alpha+\sum_{j=1}^p\beta_jx_{ji},
\end{equation}
where $\tau \in \Re$ is the tail parameter. Since a GEV link can be asymmetric, underestimation of the default probability may be overcome. As discussed, for instance, in \cite{Calabrese:2013}, depending on the value of $\tau$, several special cases can be recovered; e.g., when $\tau\rightarrow 0$ the GEV random variable follows a Gumbel distribution and its cumulative distribution is the log-log function \citep{Agresti:2002}. In this way, \cite{Calabrese:2013} propose the \emph{GEV regression model}.

Second, the logistic and the GEV (\ref{linkgev}) models assume a linear relationship between the explanatory variables and the response $\eta_i$.  These models can mask possibly interesting non-linear patterns which can help improve our understanding of the underlying covariate-response relationships
and perhaps improve the prediction accuracy of the scoring model as well \citep{Berg:2007, Marra:2013, Chuang:2009, Gestel:2005,Huang:2006,Lee:2005,Lin:2012,Ong:2005}. Therefore \cite{Marra:2013} propose the BGEVA model, an extension of the GEV model based on penalized regression splines to flexibly determine covariate effects from the data.

In the GEV model, the right part of equation (\ref{linkgev}) is changed to obtain an additive model given by
\begin{equation}
\label{linkbgeva}
\frac{\left[-\ln (PD_i)\right]^{-\tau}-1}{\tau}=\alpha+\sum_{j=1}^p\beta_js(x_{ji}),
\end{equation}
where the $s_j(x_{ij})$ are unknown one-dimensional smooth functions of the continuous covariates $x_{ji}$.

The smooth functions $s(x_{ij})$ in the model are approximated
by a linear combination of $K_j$ known (e.g., cubic or thin plate regression) spline bases, $b_k(x_{ji})$ and unknown regression parameters, $\gamma_{jk}$ \citep{Wood:2006,Marra:2013}:
$$s_j(x_{ji})=\sum_{k=1}^{K_j} \gamma_{jk} b_k(x_{ji}).$$

Calculating $b_k(x_{ji})$ for $k$ and each observation point gives $K_j$ curves with different degrees of complexity which multiplied by some real valued parameters $\gamma_{jk}$ and then summed to give an estimated curve for the smooth component \citep{Ruppert:2003}.
Replacing in model (\ref{linkbgeva}) the smooth terms with their regression spline expressions yields essentially a classic parametric model.
Estimating the $\beta_j$ parameters and the smooth functions $s(x_{ij})$ we can predict the default probabilities by using the inverse of the equation (\ref{linkbgeva}).
The smooth functions show the existence of possible non-linear relationships between the response variable and the predictors and allow us to improve on the prediction results obtained using classic alternatives.
The model is implemented in the \texttt{R} package \textit{bgeva} \citep{bgeva} available for download from CRAN.

SMEs may not provide full details of their financial statements \citep{Sohn:2013,Ciampi:2013}, for this reason missing values could be a problem for scoring models for SMEs \citep{Lin:2012, Ciampi:2013}. A widely used method for missing values is multiple imputation, which was proposed by \cite{Rubin:1977} and described in detail by \cite{Graham:2012}. Multiple imputation can be described as a three-step process. First, sets of plausible values for missing observations are created. Each of these sets of plausible values can be used to 'fill-in' the missing values and create a 'completed' dataset. Second, each of these datasets can be analysed using complete-data methods. Finally, the results are combined, which allows the uncertainty regarding the imputation to be taken into account.

In this paper we use an MCMC algorithm known as fully conditional specification \cite{Graham:2012}. The basic idea is to impute incomplete variables one at time by liner regression, using the filled-in variable from one step as a predictor in all subsequent steps. Few works apply multiple imputation to credit scoring models \citep{Cosh:1999, Crook:1996, Twala:2009, Lopez:2010}. The latter study found MCMC multiple imputation to be superior to other methods of handling missing values, therefore, it is used in this research.

Another approach to cope with missing values is based on so-called coarse-classification \citep{Thomas:2002}. This procedure consists in dividing the values of a numeric predictor into categories or classes. Normally there are 10-20 fine classes initially produced for the range of ordered values from minimum to maximum. In this paper we divide the numeric predictors into 10 classes of approximately the same size (maintaining exactly the same size is not possible because of the varying numbers of missing values for different variables).

For each fine class a proportion of defaults (or bad accounts or simply Bads) is calculated, and adjacent categories can be further grouped together into coarse classes, if the default rates are sufficiently close. Missing values are entered as a separate category. Categories can be entered into the model as binary dummies or alternatively are transformed into Weights of Evidence (WoE):
\begin{equation} \label{WOE}
 WoE_i=ln\left[\frac{b_i/g_i}{B/G}\right]=ln\left(\frac{b_i \:G}{g_i\:B}\right),
 \end{equation}
where $b_i$  is the number of bads (defaults) in category $i$ of a variable,
$g_i$  is the number of goods (non-defaults) in category $i$,
$B$ is the total number of Bads,
$G$ is the total number of Goods in the sample.

Given the fact that logistic regression is the most commonly used approach in credit scoring \citep{Thomas:2002}, WoE is appealing since this transformation produces log odds measures (same scale as logistic regression). Furthermore, log-odds of each category are compared to that of the sample: positive values would indicate riskier classes and negative values - more creditworthy customers.

\cite{Crook:2007} report this methodology as the one most widely used within retail banking. \cite{Lin:2012} have applied it to small business distress modelling and found that it improved the predictive accuracy of the models. We use this approach as the benchmark to compare the performance of alternative methods to cope with missing values (multiple imputation) and non-linearity (BGEVA model).

\section{Empirical Analysis}
\subsection{Data description}
The empirical analysis is based on explanatory variables from 2010 to predict the default in 2011 for $39,785$ UK SMEs and $154,934$ Italian SMEs. The data are from AMADEUS-Bureau van Dijk (BvD), a database of comparable financial and business information on Europe's public and private companies. The time horizon considered here is of extreme interest as it includes the European sovereign debt crisis of 2011. In summer 2011 interest rates on Italian national debt went out of control.

The definition of SME by the European Commission is adopted. That is, a business must have an annual turnover of less than 50 million of Euro, a balance sheet total less than 43 million of Euro and the number of employees should not exceed 250 (\href{http://ec.europa.eu/enterprise/\\policies/sme/facts-figures-analysis/sme-definition/index.htm}{http://ec.europa.eu/enterprise/\\policies/sme/facts-figures-analysis/sme-definition/index.htm}).
Furthermore, the number of subsisdiaries is capped at 6, in accordance with \cite{Lu:2001}, and the number of directors is  10 maximum, consistent with \cite{Gabrielsson:2007,Michala:2013}.

In this work, we consider a default to have occurred when a specific SME enters a bankruptcy or a liquidation procedure. Moreover, a SME is classified as default also if it is active and it has not paid a debt (classified as default of payment by BvD) or it is in administration or receivership or under a scheme of arrangement (defined as insolvency proceedings by BvD). On the contrary, non-defaulters include active and dormant SMEs (only 29 for both samples). A dormant company is still registered, but has no significant activity (and no significant accounting transactions during the accounting period). Consistent with previous studies \citep{Altman:2007,Altman:2010, Pederzoli:2013} we exclude dissolved firms that no longer exist as a legal entity, but the reason for this is not specified. This is in line with the objective of this paper that  models the probability of going bankrupt using publicly available information.  Dissolved category comprises SMEs that may not necessarily experience financial difficulties, they may stop trading because the owner retires or for similar reasons. Future research can investigate dissolved as a separate category.

The use of the common database has ensured the availability of the common set of variables measured in the same way for both countries. We used financial ratios that have been found important in previous research on SMEs \citep{Altman:2007,Lin:2012,Michala:2013}. Adopting the classification of variables suggested in \cite{Altman:2007} the variables in this research covered all five major groups usually used:

\begin{itemize}
\item Leverage (e.g. Gearing, Solvency ratio);
\item Liquidity (e.g. Current ratio, Liquidity ratio, Shareholder liquidity ratio);
\item Profitability (e.g. EBITDA margin, Profit Margin, ROCE, ROE);
\item Coverage (e.g. Interest cover);
\item Activity /Scale/Size (e.g. Total assets, Shareholder funds, No of employees, No of directors, No of subsidiaries).
\end{itemize}

Following \cite{Michala:2013} who found cash flow management significant in predicting default, we also include cash flow based measures (e.g. Cash flow, Cash flow / Operating revenue).  The variables have been checked for linear dependence, and highly collinear ones have not been used in the analysis.
Table 1 presents short and full names of the variables initially considered and some descriptive statistics on the training sample.
\\
\\
Table 1 around here.
\\
\\
The SMEs in the UK sample are larger as compared to Italian SMEs in terms of Total assets, Operating revenue, No of employees, No of directors. This is consistent with the EU statistics reported in Sections 1-2. The summary statistics for Age and No of  subsidiaries are similar for the two countries. The UK businesses have higher liabilities, but profitability is also higher. The Italian companies show better Cash flow and lower debt.
Despite using the common source of the data, the percentages of missing values are different across the countries. For Italy, the variable with the highest number of missing is Cash flow / Operating revenue, with 19.5\% missing. For the UK, the problem is much more acute, the highest percentage of missing is 59.2\% for ROCE.  This has an effect on the results, depending on how missing values have been treated, as can be seen from Tables 2 and Table 3 that show the variables that are significant at 10\% level or lower across the models.
\\
\\
Table 2 around here
\\
\\
Table 3 around here

\subsection{Predictive accuracy}

To avoid sample dependency, the predictive accuracy for the models was tested on control samples, i.e. we used out-of-sample tests. For each country the whole dataset was split into training (70\%) and control (30\%) samples using a stratified random sampling with stratification on default indicator. Measures of predictive accuracy used include mean absolute error (MAE), mean square error (MSE) and Area under the ROC curve (AUC). MAE and MSE are standard measures of predictive accuracy in forecasting studies. Obviously, scoring models with lower MSE and MAE should forecast defaults and non-defaults more accurately. For a bank it is much more costly to classify an SME as a non-defaulter when it is a defaulter than the opposite.
If a defaulter is classified as a non-defaulter, then it will be accepted for credit,  which will subsequently be lost (in part or as a whole). Yet when a non-defaulter is classified as a defaulter, it is only a lost opportunity. Therefore, in this study MSE and MAE are reported for defaults only and they are denoted by MSE$^+$ and MAE$^+$. AUC is the most popular measure of model performance in credit scoring \citep{Thomas:2002} that summarises the ability of the model to rank-order the risk correctly over the whole range of predicted PDs. Higher value indicate better performance.
\\
\\
Table 4 around here
\\
\\
Table 5 around here
\\
\\
Tables 4 and 5 summarise the results\footnote{To obtain these results we use SPSS for imputed missing values and the package ''bgeva" of R-program.} for the UK and Italian models for imputed and Weights of Evidence (WoE) data.

Considering WoE approach on the UK data, the GEV model shows better performance on both error measures and AUC than the logistic model (Table 4).
This can be attributed to GEV overcoming the problem of the PD underestimation by the logistic model. Moreover, by applying the non-parametric model (BGEVA) the performance on  MSE$^+$ and MAE$^+$ improves further. This fact justifies the use of a non-parametric credit scoring model that can capture non-linear relationships between the accounting characteristics of SMEs and response.

As for imputed values on the UK data, the best MAE$^+$ and MSE$^+$  are for BGEVA, whilst the best AUC is shared between BGEVA and additive logistic model. This further emphasises the advantage of BGEVA \textbf{in forecasting defaults in low default portfolios} that performs well on both methods of treating the missing values.

Considering WoE approach on Italian data (Table 5), we observe results  similar to the UK models. BGEVA has the best MAE$^+$ and MSE$^+$, whilst additive logistic produces slightly higher AUC, but the difference is negligible. For Italian imputed values the results are mixed. The additive logistic model shows the lowest values of the MAE$^+$ and MSE$^+$, whilst
the GEV and logistic models show higher values of the AUC.

The comparison of the predictive accuracy between the countries should be interpreted with caution due to the different sample sizes, different proportions of missing values and different number of significant variables (as discussed in the next section). Since the UK sample size is smaller than the Italian one and the percentage of UK missing values is higher than for Italy (see Table 1), one can expect a decrease in the predictive accuracy.
However, for completeness it could be stated that
all models for Italy have better performance than the UK models. Moreover, the Italian best model (BGEVA) has also a lowest MAE$^+$.

It should also be noted that WoE coding provides better performance as compared to Imputation with the only exception of MAE$^+$ of BGEVA for the UK. An explanation for this may be a non-random nature of missing values.

In conclusion, the empirical results confirm that the BGEVA model performs well for SMEs default forecasting for both countries. This can be attributed to the fact that the linearity assumption is not  supported by the data of both countries, as will be discussed in the next section.

\subsection{Comparison of risk predictors between Italian and UK SMEs}

There are differences between the countries in terms of significant variables and their number depending on the model/approach used. Whilst logistic regression for both countries and GEV model for Italy show the same number of variables irrespective of imputation or WoE, there are differences in model composition even in these cases. For example, in logistic regression for the UK - Cash flow, Interest cover and Operating revenue are significant with WoE coding, but not with Imputation; yet with Imputation the following variables become significant: Profit margin, Shareholder funds and Total assets. For the rest of models the numbers of significant variables differ with the extreme cases of GEV and BGEVA for the UK, where WoE coding increases the number of significant variables from 11 to 20. This may be interpreted as suggesting that at least for some variables values cannot be assumed to be missing at random, therefore WoE increase the number of significant variables.

Only two variables consistently appear across all 16 models for the two countries:  No of directors  and Solvency ratio (Tables 2 and 3). No of subsidiaries appear in all models, but one. Profit margin and Shareholder funds enter 14 models. Other frequent variables that are significant at 10 per cent level or lower across all 16 models for the two countries are Liquidity ratio (13), Age (12), EBITDA margin (12), No of employees (12), Operating revenue (12), Cash flow / Operating revenue (10), Total assets (10), ROE (10). When looking at most frequent significant variables for each country separately (e.g. common variables that are in more than half of the models for each country) these  include No of directors, Solvency ratio, No of subsidiaries, Profit margin, Shareholder funds and Liquidity ratio. This confirms the results from previous research that suggests measures of profitability, leverage and liquidity are important \citep {Altman:1997, Altman:2010,Michala:2013}. Shareholder funds can be interpreted as the interest the shareholders have in the company, and also the ability of the company to raise funds for growth/expansion. Solvency ratio emphasis the importance of the proportion of Shareholder funds in the assets of the company. No of directors and No of subsidiaries may be interpreted as proxies for company size and the scale of the activity, with No of directors also acting as a crude proxy for quality of management (assuming more directors would mean better management).
\\
\\
Table 6 around here
\\
\\
Table 7 around here
\\
\\
Despite the commonality reported above, there are some interesting differences between the countries. The most notable one is the fact that Gearing is significant in all UK models, whilst not being significant in Italy at all. This suggests the importance of the firm's ability to pay both long-term debt and short-term one in the UK. For Italy measures of profitability are relatively more prominent: EBITDA margin and ROE appear in almost all Italian models, in addition to Profit margin which is common to both countries.
Age and No of employees are twice more frequent in the UK models. Age has been previously found important in  \cite{Altman:2007}. No of employees indicates the size of the company or its scale. Financial scale for Italy is most frequently represented by Operating revenue, which appears in all Italian models, but only in half of the UK ones. Cash flow/ Operating revenue is also present in all Italian models.

As an example of more detailed cross-country differences, consider the estimates of BGEVA model on imputed values presented in Tables 6. The interpretation of WoE is less straightforward since it requires the information on category boundaries and WoE values. This information and details of other models are available on request.
Financial measures common to both countries include ratios of profitability (Profit margin), leverage (Solvency ratio), liquidity (Liquidity ratio) and scale (Shareholder funds, Total  assets). In addition, there are common non-financial variables across the two countries: Age, No of directors,  No of employees, No of subsidiaries. This fact emphasises the value of non-financial information in modelling SMEs and confirms some previous research \citep{Altman:2010}.

Tables 6 and 7 report the estimation results of the parametric and non-parametric components of the BGEVA model for the two countries and for multiple imputation. Some of the covariate effects are reported in the parametric part of the BGEVA model
since their smooth function estimates were linear. Explanatory variables significant deviations from the linearity assumption are reported in the smooth terms part.
The variables show different degrees of non-linearity (Edf). The parameter Edf (degrees of freedom) in Tables 6 and 7 controls the smoothness of the curve. The variables with Edfs equal to 1 show linear smooth function so they are reported in the parametric part. The estimated smooths that exhibit Edfs considerably greater than 1 are reported in smooth terms part. Larger Edf allows a very flexible curve, e.g., a curve that can have multiple local maxima and minima. The values of degrees of freedom are estimated from the data.  The most interesting smooth terms are displayed in the Figures 1 and 2. In line with the interpretation for the parametric components, if the estimated smooth function of a covariate is decreasing then the estimated $PD$ decreases when the explanatory variable increases, and vice versa.
\\
\\

Figure 1 around here
\\
\\
Figure 2 around here
\\
\\
There is some commonality between the countries with Liquidity ratio and Age being non-linear for both countries.  No of directors and Total assets exhibit non-linear relationship with the response for the UK, but not for Italy, on the contrary, Cash Flow and  No of employees show non-linear patterns for Utaly only.

Consider Liquidity ratio that shows a non-linear relationship for both countries (Figures 1 and 2). For Italy when this variables increases, the PD decreases (although in a non-linear way), in accordance with the expectations and prior research by \cite{Pederzoli:2013}. Yet for the UK the relationship is more complex. Up to 30 and from 75 the relationship of this covariate  to PD is negative (as expected). However, in the middle section it is the opposite: increasing values of Liquidity ratio signal increasing  chances of default. This may be related to difficulties in getting  credit for SMEs, if Current Liabilities in denominator are decreasing.

Previous research summarised in Section 2 did not use exactly the same ratio, yet \cite{Altman:2010} report a negative relationship between a similar variable (Current ratio) and the PD. It should be noted though, that the authors did not comment on potential non-linearity. For German SMEs \cite{Figini:2009} and \cite{Figini:2012} observed a counter-intuitive sign for Liabilities ratio and explained it by the fact that many small business owners cover their debts from external sources.

Examples of variables that show non-linear relationships and are not common for the two countries are Total Assets for the UK and No of Employees for Italy, both  can be interpreted as proxies for SME size. From Figure 2 looking at Total assets we can deduce that the UK small and micro enterprises show higher default risk, in line with \cite{Figini:2009} for German SMEs. Then for companies with Total assets higher than 20 million euros, when this variable increases the PD decreases.\cite{Altman:2010} also noted the non-linear nature of Total assets. Finally, from the plot for Number of Employees (Figure 1) Italian small and micro enterprises have higher PD when the number of employees increases. For medium enterprises this relationship becomes negative, although the confidence intervals are wide.
These results highlight some interesting patterns observed from the data, yet further research would be beneficial in order to fully understand the implied relations.

\section{Conclusions and extensions}
This paper has compared predictors of SMEs insolvencies across the UK and Italy, using publicly available information from 2010 to model the company status in 2011. The choice of the time period after the credit crisis makes this comparison particularly relevant, due to different economic situations in the two countries. Whilst Italy was experiencing high interest rates for its national debt, that was not the case in the UK despite the latter showing low economic growth. There are also differences across the two countries in the relative importance that SME play in the two economies, as discussed in Section 2. Despite these differences, there were some financial measures significant in predicting insolvency. These included measures of profitability, leverage, liquidity and scale. In addition, there was some commonality in non-financial measures, thus highlighting the importance of soft information for analysis of SME performance. As for the differences, profitability measures are significant more frequently for Italy, whilst for the UK Gearing is a significant predictor, not featuring in Italian models.

A number of different modelling approaches have been explored in order to improve predictive accuracy. Generalised Extreme Value (GEV) regression was applied to correct for the symmetric link function of the logistic regression, which is a standard approach in credit risk modelling.  The assumption of non-linearity was relaxed through application of BGEVA, non-parametric additive model based on the GEV link function. In addition, two methods of handling missing values were compared: multiple imputation and Weights of Evidence (WoE) transformation. The results suggest that the best predictive performance is obtained by BGEVA, thus implying the necessity of taking into account the relative volume of defaults and non-linear patterns when modelling SME insolvencies. WoE generally showed better prediction as compared to imputation, suggesting that missing values are informative and cannot be assume to be missing at random.

This study presents an initial attempt to understand the cross-country drivers of SMEs insolvencies, and is exploratory in the general approach adopted. Further extensions could include exploration of additional countries and additional variables, in particular, of non-financial nature, but this depends on the data availability. Causal relations through structural equation models can be investigated. On the practical side, it would be of interest to consider predictors significant to both countries and construct a generic model with the objective of comparing it to a country-specific model. Finally, different groups of SMEs that go out of business can be explored, e.g. dissolved.
\begin{figure}[htbp]
	\centering
		\includegraphics[width=1\textwidth]{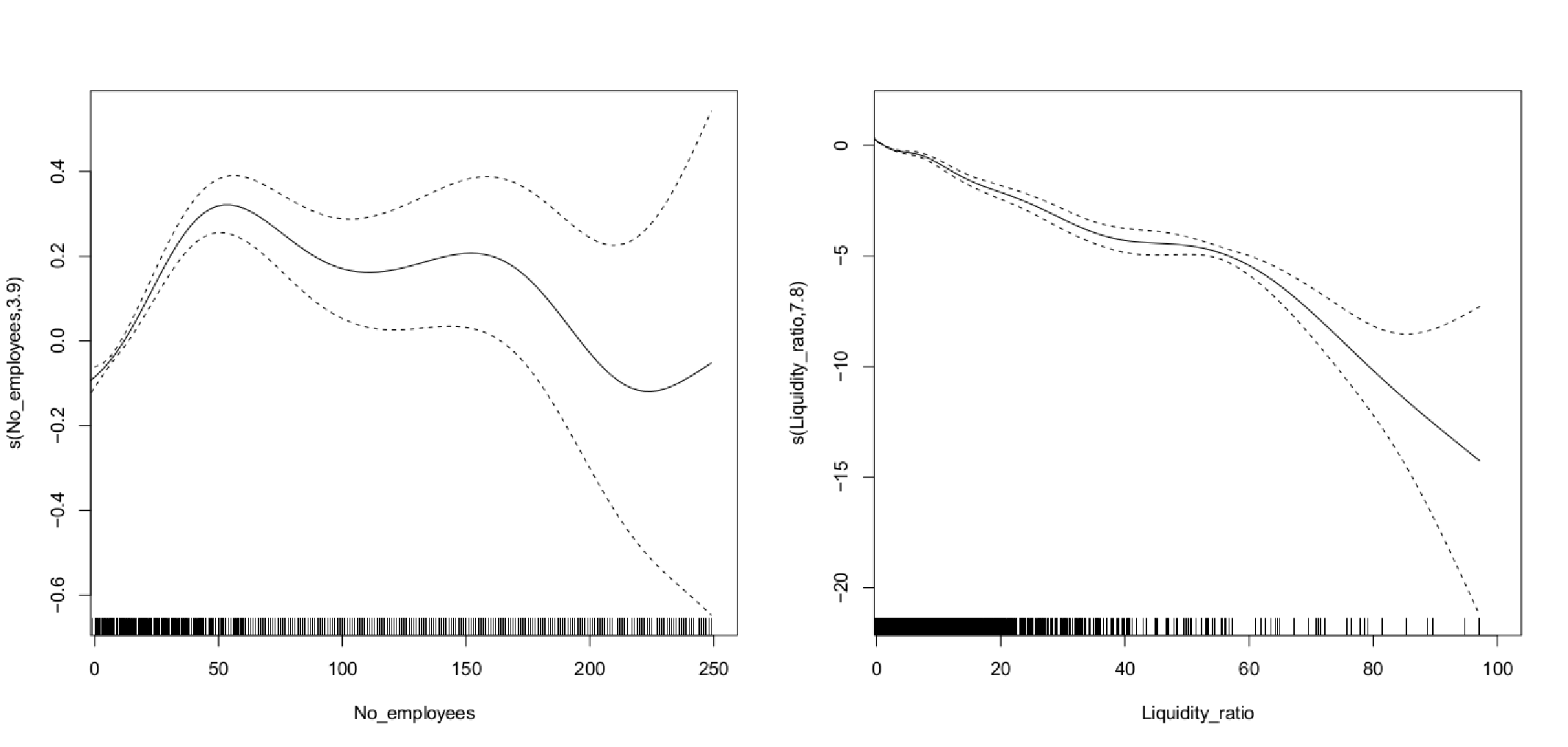}
\caption{Smooth component estimates of the 2 (out of 4) continuous variables that exhibit a non-linear pattern. These were obtained from applying the BGEVA model on the Italian SME data. Results are on the scale of the predictor. The plot show the $95\%$ confidence intervals. The numbers in brackets in the y-axis captions are the estimated degrees of freedom (Edf) of the smooth curves.}
\end{figure}
\begin{figure}[htbp]
	\centering
		\includegraphics[width=1\textwidth]{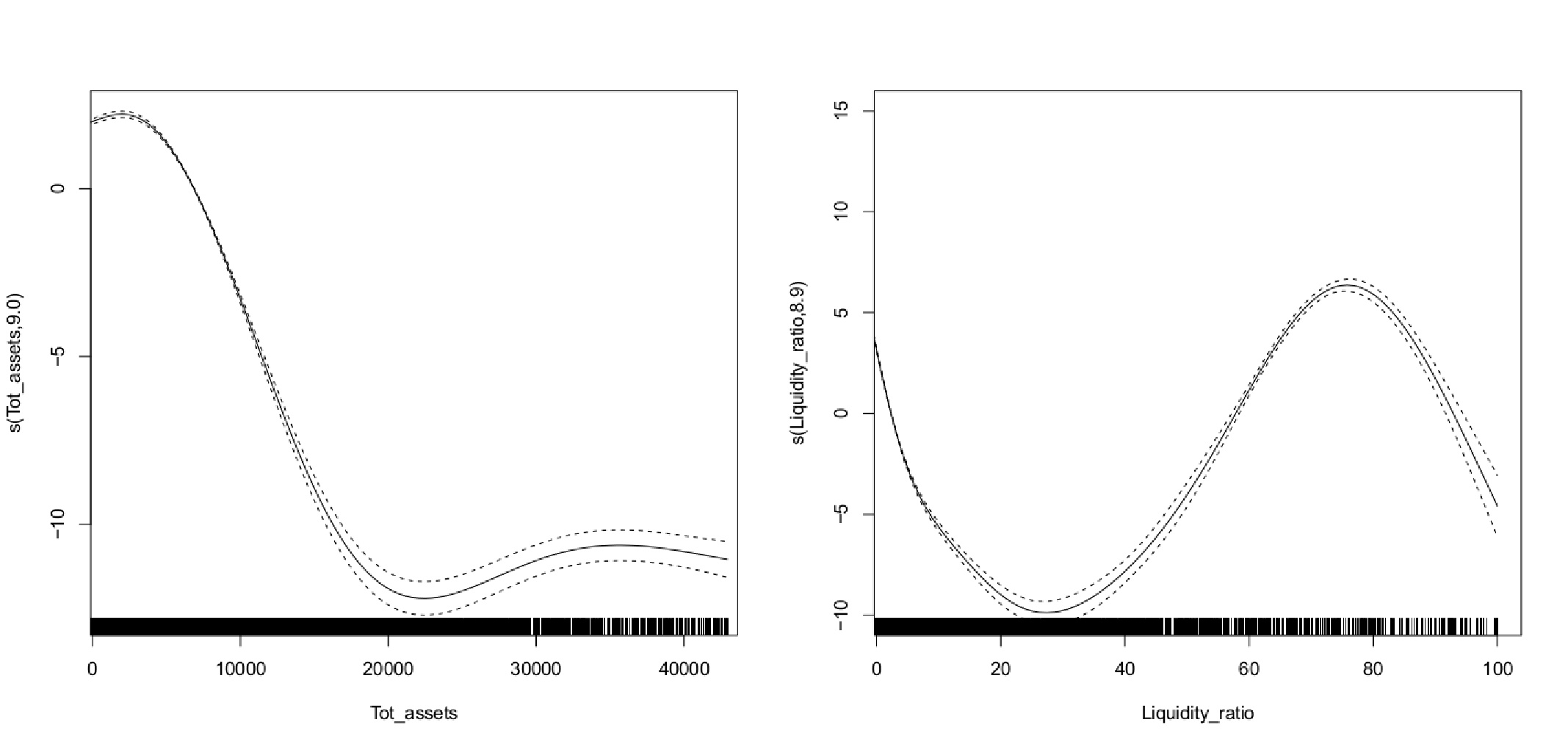}
\caption{Smooth component estimates of the 2 (out of 4) continuous variables that exhibit a non-linear pattern. These were obtained from applying the BGEVA model on the Italian SME data. Results are on the scale of the predictor. The plot show the $95\%$ confidence intervals. The numbers in brackets in the y-axis captions are the estimated degrees of freedom (Edf) of the smooth curves.}
\end{figure}

\begin{landscape}
\begin{table}\label{descriptive1}
\begin{center}
\begin{tabular}{|l|l|c|c|c|c|c|c|}
\hline
\multicolumn{2}{|c|}{\emph{Variables}}  &  \multicolumn{3}{|c|}{\emph{Italy, n=106967}}& \multicolumn{3}{|c|}{\emph{UK, n=27132}}\\
\emph{Short name} & \emph{Description} &  \emph{\% Missing} & \emph{Mean} & \emph{Std Dev}&  \emph{\% Missing} & \emph{Mean} & \emph{Std Dev}\\
\hline
Age		          & Age of the company, months			              & 0.2 	& 207.55 &	156.072		& 0.2	& 206.59  &	192.983\\
Capital		      & Capital, th EUR				              & 0.2	    & 203.46 &	816.490		& 2.0	& 331.96  &	2689.960\\
Cash$\_$flow		          & Cash flow, th EUR			              & 2.4	    & 109.23 &	617.864		& 21.4	& 47.69	  & 40224.523\\
Cash$\_$flow$\_$oprev	      &	Cash flow / Operating revenue,\%          & 19.5	& 7.00	 &  8.250		& 39.4	& 11.98	  & 14.668\\
Current$\_$liab          &	Current liabilities, th EUR		          & 0.2	    & 1606.02&  3045.892	& 1.6	& 2368.52 & 6464.733\\
Current$\_$ratio	          &	Current assets/Current liabilities,\%             & 0.3	    & 1.76	 &  2.767		& 3.7	& 4.75	  & 10.125\\
EBITDA$\_$Margin	          &	EBITDA/Operating revenue, \%			              & 3.5	    & 6.79	 &  14.564		& 23.2	& 8.64	  & 20.992\\
Gearing		      & (Long term liab. + Short term loans)/       & 16.5	& 188.34 &  220.371& 20.7	& 76.62	  & 149.386\\
                                      & Shareholders funds, \%	 &        &           &         &            &            &            \\
Interest$\_$cover	      &	P(L) before interest/ Interest paid, \%		           & 8.2	    & 26.13	 &  96.408		& 57.6	& 39.52	  & 119.958\\
Liquidity$\_$ratio      &	 (Current assets - Stock)/Current liab.              & 0.3	    & 1.36	 &  2.392		& 5.0	& 4.53	  & 10.119\\
Loans			          & Loans, th EUR				              & 0.2	    & 437.09 &	1300.625	& 3.3	& 1025.80 & 4005.364\\
Net$\_$income		      & Net income, th EUR 	       & 0.2	    & 11.46	 & 635.865		& 3.2	&109.074 & 4357.018\\
No$\_$directors          &	Number of current directors/managers	  & 0.0	    & 1.33	 & 1.810		& 0.0	& 4.74	  & 2.555\\
No$\_$employees          &	Number of employees			              & 0.2	    & 13.52	 & 22.239		& 2.2	& 37.19	  & 47.120\\
No$\_$subsidiaries	      &	No of recorded subsidiaries		          & 0.0	    & 0.41	 & 0.872		& 0.0	& 0.47	  & 0.977\\
Noncurrent$\_$liab	      &	Non-current liabilities, th EUR	      & 0.2	    & 613.73 & 1602.776	    & 1.7	& 1001.87 &	4763.591\\
Op$\_$rev 			      & Operating revenue (Turnover), th EUR 	  & 0.2	    & 2998.49& 5457.546	    & 1.7	& 6152.03 &	8545.211\\
PL$\_$beforetax	          &	Profit (Loss) before tax, th EUR 			          & 0.2	    & 56.38	 & 606.305		& 3.0	& 179.731   &  4391.610\\
Profit$\_$employee	      &	Profit per employee, th EUR		          & 2.1	    & 10.35	 & 60.350		& 7.0	& 28.41	  & 182.253\\
Profit$\_$margin	          &	P(L) before tax/ Operating revenue, \%			              & 1.9	    &  1.03	 & 14.395		& 7.8	& 7.66	  & 25.868\\
ROCE			          & P(L) before tax/ (Total assets - Cur. liab.)		      & 6.5	    & 10.77	 & 57.921		& 59.2	& 19.08	  & 81.989\\
ROE			          & P(L) before tax/ Shareholder funds, \%		      & 7.7	    & 13.36	 & 97.496		& 18.0	& 25.85	  & 109.174\\
Shareh$\_$liquidity$\_$ratio  &	Shareholders funds/ Long term liab., \%          &	2.2	    & 7.46	 & 37.293		& 52.0	& 36.03	  & 105.713\\
Sharehold$\_$funds	  & Shareholders funds, th EUR 		          & 0.2	    & 865.00 & 2381.846	    & 1.6   & 1495.01 &	7183.142\\
Solvency$\_$ratio		  & Shareholders funds/Total assets, \%		  & 1.1	    & 23.35	 & 24.527		& 6.3	& 45.43	  & 38.256\\
Tot$\_$assets		      & Total assets, th EUR			          & 0.2 	& 3084.70& 5422.522	    & 1.8   & 4860.58 &	6705'262\\
 \hline
\end{tabular}
\caption{Descriptive statistics for training samples}
\end{center}
\end{table}
\end{landscape}
\begin{table}\label{descriptive3}
\begin{center}
\begin{tabular}{|l|c|c|c|c|c|c|c|c|c|}
\hline
\multicolumn{1}{|c|}{\emph{Variables}}  &  \multicolumn{4}{|c|}{\emph{Logistic regression}}& \multicolumn{4}{|c|}{\emph{Additive Logistic regression}}& \emph{times in } \\
    & \multicolumn{2}{|c|}{\emph{Italy}} &  \multicolumn{2}{|c|}{\emph{UK	}} & \multicolumn{2}{|c|}{\emph{Italy}} & \multicolumn{2}{|c|}{\emph{UK	}} & \emph{all 16 models} \\
\emph{Short name}     & \emph{Imp} &  \emph{Woe} & \emph{Imp} & \emph{Woe}&  \emph{Imp} &  \emph{ Woe} & \emph{Imp} & \emph{Woe	}&\\
\hline
Age			              & X &	0 &	X &	X &	SX &  0 &	SX &  X & 12\\
Cash$\_$ flow			  & X &	0 &	0 &	X &	SX &  0 &	0  & SX	& 8\\
Cash$\_$flow$\_$oprev	  & X &	X & 0 & 0 & X  &  X &   0  & 0	& 10\\
Current$\_$ratio	          &	0 & 0 & 0 & 0 & X  &  0	&   0  & 0	& 2\\
EBITDA$\_$Margin	          &	X & X &	0 &	0 &	X  &  X	&   0  & X  & 12\\
Gearing                   &	0 &	0 & X & X &	0  &  0	&   X  & X  & 8\\
Interest$\_$cover	      &	0 & X &	0 &	X & 0  &  SX&	0  & X  &  8\\
Liquidity$\_$ratio		  & X &	X &	0 &	X &	SX &  0 &	SX & X  & 13\\
Net$\_$income	          &	0 &	X &	0 &	0 &	0  & SX	&   0  & 0  & 6\\
No$\_$directors	          &	X &	X &	X &	X &	X  & X	&  SX  & X  & 16\\
No$\_$employees			  & X &	0 &	X &	X &	SX & 0	&   X  & X  & 12\\
No$\_$subsidiaries		  & X &	X & X &	X &	 X & X	&   0  & X  & 15\\
Op$\_$rev	              &	X &	X &	0 &	X &	 X & X	&   0  & SX & 12\\
PL$\_$beforetax	          &	0 &	X &	0 & 0 &	 0 & SX	&   0  &  X & 7\\
Profit$\_$margin           &	X &	X &	X &	0 &	 X &  X	&   X  &  0 & 14\\
ROCE	                  &	0 &	X &	0 &	0 &	 0 & SX &   0  &  0 &  6\\
ROE                  &	X &	X &	0 &	0 &  X &  0	&   0  & SX & 9\\
Sharehold$\_$funds		  & X & X &	X &	0 &	 0 &  0	&   X  &  0 & 14\\
Shareh$\_$liquidity$\_$ratio  & 0 &	0 &	0 &	0 &	X  &  X &	X  &  0 & 5\\
Solvency$\_$ratio         &	X &	X &	X &	X &	X  &  X	&   X  &  X	& 16\\
Tot$\_$assets		      & X &	0 &	X & 0 & X  &  0	&  SX  &  0 & 10\\
 \hline
\end{tabular}
\caption{Significant variables across the countries for logistic and additive logistic models. X - the variable is significant at 10\% s.l. or lower; SX - the smooth term of the variable is significant at 10\% s.l. or lower}
\end{center}
\end{table}
\begin{table}\label{descriptive4}
\begin{center}
\begin{tabular}{|l|c|c|c|c|c|c|c|c|c|}
\hline
\multicolumn{1}{|c|}{\emph{Variables}}  &  \multicolumn{4}{|c|}{\emph{Gev model}}& \multicolumn{4}{|c|}{\emph{BGEVA model}}& \emph{times in} \\
    & \multicolumn{2}{|c|}{\emph{Italy}} &  \multicolumn{2}{|c|}{\emph{UK	}} & \multicolumn{2}{|c|}{\emph{Italy}} & \multicolumn{2}{|c|}{\emph{UK	}}&   \emph{all 16 models} \\
\emph{Short name}     & \emph{Imp} &  \emph{Woe} & \emph{Imp} & \emph{Woe}&  \emph{Imp} &  \emph{ Woe} & \emph{Imp} & \emph{Woe	}&\\

\hline
Age			          & X &	0 &	X &	X &	SX &  0 &	SX &  X & 12\\
Cash$\_$flow			      & X &	0 &	0 &	X &	SX &  0 &	0  & SX	& 8 \\
Cash flow$\_$oprev		  & X &	X & 0 & X & X  &  X &   0  & X	& 10\\
Current$\_$ratio	          &	0 & 0 & 0 & 0 & X  &  0	&   0  & 0	& 2 \\
EBITDA$\_$Margin	          &	X & X &	0 &	X &	X  &  X	&   0  & X  & 12\\
Gearing              &	0 &	0 & X & X &	0  &  0	&   X  & X  & 8 \\
Interest$\_$cover	      &	0 & X &	0 &	X & 0  &  SX&	0  & X  &  8\\
Liquidity$\_$ratio		  & X &	X &	X &	X &	SX &  0 &	SX & X  & 13\\
Net$\_$income	          &	0 &	X &	0 &	X &	0  & SX	&   SX  & X  & 6\\
No$\_$directors	          &	X &	X &	X &	X &	X  & X	&  SX  & X  & 16\\
No$\_$employees			  & X &	0 &	X &	X &	SX & 0	&   X  & X  & 12\\
No$\_$subsidiaries		  & X &	X & X &	X &	 X & SX	&   X  & X  & 15\\
Op$\_$rev	              &	X &	X &	0 &	X &	 X & X	&   0  & SX & 12\\
PL$\_$beforetax	          &	0 &	X &	0 & X &	 0 & 0	&   SX  & 0 & 7 \\
Profit$\_$margin           &	X &	X &	X &	X &	 X &  X	&   X  &  0 & 14\\
ROCE	                  &	0 &	X &	0 &	X &	 0 & SX &   0  &  X &  6\\
ROE                 &	X &	X &	0 &	0 &  X &  0	&   0  & SX & 9 \\
Sharehold$\_$funds		  & X & X &	X &	X &	 X &  X	&   X  &  X & 14 \\
Shareh$\_$liquidity$\_$ratio  & 0 &	0 &	X &	X &	0  &  0 &	X  & SX & 5\\
Solvency$\_$ratio        &	X &	X &	X &	X &	X  &  X	&   X  &  X	& 16\\
Tot$\_$assets	      & X &	0 &	X & X & X  &  0	&  SX  &  X & 10\\
 \hline
\end{tabular}
\caption{Significant variables across the countries for Gev and BGEVA models. X - the variable is significant at 10\% s.l. or lower; SX - the smooth term of the variable is significant at 10\% s.l. or lower}
\end{center}
\end{table}

\begin{table}
\label{UKaccuracy}
\begin{center}
\begin{tabular}{|l|c|c|c|c|c|}
\hline

\emph{Methods for missing values} & \hspace{1mm}\emph{measure} & \emph{GEV model} & \emph{logistic}& \emph{BGEVA model} & \emph{additive logistic}\\
\hline

  \multirow{3}{*}{\vspace{8mm}\emph{Weight of Evidence }}          &  MAE$^+$      & 0.784   & 0.798 & 0.782 & 0.797 \\
                                                                   &  MSE$^+$     & 0.722   & 0.705  & 0.702 & 0.702 \\
                                                                   &  AUC         & 0.741   &  0.731 & 0.722 &  0.717\\
   \hline
  \multirow{3}{*}{\vspace{8mm}\emph{Imputation}}                   &  MAE$^+$      & 0.862  & 0.909 & 0.761 & 0.969\\
                                                                   &  MSE$^+$     & 0.807  & 0.838 & 0.713 & 0.941\\
                                                                   &  AUC         & 0.632  & 0.632  & 0.677 & 0.677\\
                                                                  \hline
    \end{tabular}
\caption{Forecasting accuracy measures for out-of-sample exercise obtained from applying the Gev and logistic model and BGEVA and logistic additive models to Uk data.}
\end{center}
\end{table}

\begin{table}
\label{Itaccuracy}
\begin{center}
\begin{tabular}{|l|c|c|c|c|c|}

\hline

\emph{Methods for missing values} & \hspace{1mm}\emph{measure} & \emph{GEV model} & \emph{logistic}& \emph{BGEVA model} & \emph{additive logistic }\\
\hline

  \multirow{3}{*}{\vspace{8mm}\emph{Weight of Evidence }}          & MAE$^+$      & 0.803   & 0.804 & 0.781 & 0.782\\
                                                                   &  MSE$^+$     & 0.679   & 0.684 & 0.651 & 0.662\\
                                                                   &  AUC         & 0.813   & 0.812 & 0.824 & 0.825\\
   \hline
  \multirow{3}{*}{\vspace{8mm}\emph{Imputation}}                   & MAE$^+$      & 0.835    & 0.814 & 0.891 & 0.803 \\
                                                                   &  MSE$^+$     & 0.730    & 0.711 & 0.835 & 0.701\\
                                                                   &  AUC         & 0.806    & 0.806 & 0.799 & 0.801\\
                                                                  \hline
    \end{tabular}
\caption{Forecasting accuracy measures for out-of-sample exercise obtained from applying the Gev and logistic model and BGEVA and logistic additive models to Italian data.}

\end{center}
\end{table}

\begin{table}\label{ITUKimp}
\begin{center}
\begin{tabular}{|l|c|c|c|c|c|c|}
\hline
\emph{Variables names} & \multicolumn{3}{|c|}{\emph{Italy}}  & \multicolumn{3}{|c|}{\emph{UK}}\\
  \hline
\emph{of parametric model} & \emph{Estimate} &  \emph{Std.Error} & \emph{p-value} & \emph{Estimate} &  \emph{Std.Error} & \emph{p-value}\\
  \hline
Intercept           & -1.308e+00  & 1.526e-02   & $<$ 2e-16 & 3.888e+00  & 6.149e-02   & $<$ 2e-16 \\
Cash$\_$flow$\_$oprev & 4.011e-03& 1.308e-03 & 0.002 & - & - & - \\
Current$\_$ratio & 1.274e-01 & 1.473e-03 & $<$ 2e-16 & - & - & - \\
EBITDA$\_$Margin & -4.701e-03 & 1.153e-03 & 4.56e-05 & - & - & - \\
Gearing & - & - & - & 1.086e-02 & 2.316e-04 & $<$ 2e-16\\
No$\_$directors & -2.935e-01& 8.068e-03 & $<$2e-16 & - & - & - \\
No$\_$employees & - & - & - & 7.436e-02 & 1.512e-03 & $<$ 2e-16\\
No$\_$subsidiaries & -1.145e-01 & 1.150e-02 & $<$ 2e-16 & -9.365e-01 & 1.929e-02 & $<$ 2e-16\\
Op$\_$rev & 1.563e-05 & 2.519e-06 & 5.39e-10 & - & - & - \\
Profit$\_$margin & -3.655e-03 & 8.980e-04 & 4.70e-05 & -7.075e-02 & 1.436e-03 & $<$ 2e-16\\
ROE & -2.986e-04 & 5.703e-05 & 1.64e-07 & - & - & - \\
Shareh$\_$liquidity$\_$ratio & - & - & - & -1.416e-02 & 2.864e-04 & $<$ 2e-16\\
Sharehold$\_$funds & -1.137e-04 & 8.391e-06 & $<$ 2e-16 & 8.769e-04 & 1.811e-05 & $<$ 2e-16 \\
Solvency$\_$ratio & -8.894e-03 & 3.854e-04 & $<$ 2e-16 & -8.453e-02 & 1.724e-03 & $<$ 2e-16\\
Tot$\_$assets & 4.043e-05 & 2.595e-06 & $<$ 2e-16 & - & - & - \\
\hline
\emph{of Smooth terms} & \emph{Edf} &  \emph{Est.rank} & \emph{p-value} & \emph{Edf} &  \emph{Est.rank} & \emph{p-value}\\
  \hline
age & 2.987 &3 & 0.021 & 9.000 & 9 & $<$ 2e-16\\
Cash$\_$flow & 8.950 &9 & $<$2e-16 & - & - & - \\
Liquidity$\_$ratio & 8.084 &9 & $<$2e-16 & 8.914 & 9 & $<$ 2e-16\\
No$\_$directors & - & - & - & 9.000 & 9 & $<$ 2e-16\\
No$\_$employees & 3.898 &4 & $<$2e-16 & - & - & - \\
Tot$\_$assets & - & - & - & 9.000 & 9 & $<$ 2e-16\\
\hline
\end{tabular}
\caption{Parametric and smooth component summaries obtained from applying the semiparametric BGEVA model to the samples of Italian and Uk SMEs. The missing values are analysed by imputation method. The values of $\tau$ parameters for Italian and Uk models are $-0.41$ and $-0.9$, respectively.}
\end{center}
\end{table}
\begin{table}\label{ITUKwoe}
\begin{center}
\begin{tabular}{|l|c|c|c|c|c|c|}
\hline
\emph{Variables names} & \multicolumn{3}{|c|}{\emph{Italy}}  & \multicolumn{3}{|c|}{\emph{UK}}\\
  \hline
\emph{of parametric model} & \emph{Estimate} &  \emph{Std.Error} & \emph{p-value}& \emph{Estimate} &  \emph{Std.Error} & \emph{p-value}\\
  \hline
Intercept                  & -1.334 & 0.011 &  $<$ 2e-16 & -1.572 & 0.028  & -$<$2e-16\\
age$\_$w &- & - & - & 5.367 & 0.049 & $<$2e-16 \\
cash$\_$flow$\_$oprev$\_$w & 0.128 & 0.018 & 3.77e-12 & 0.398 & 0.021 & $<$2e-16 \\
EBITDA$\_$Margin$\_$w & 0.108 & 0.020 & 3.92e-08 & 1.506 & 0.017 & $<$2e-16 \\
Gearing$\_$w &- & - & - & -1.617 & 0.025 & $<$2e-16 \\
Interest$\_$cover$\_$w & - & - & - & 0.947 & 0.009 & $<$2e-16 \\
Liquidity$\_$ratio$\_$w & - & - & - & 0.879 & 0.013 & $<$2e-16 \\
Net$\_$income$\_$w & - & - & - & -0.536 & 0.040 & $<$2e-16 \\
No$\_$directors$\_$w & 0.588 & 0.013 & $<$2e-16 & 3.486 & 0.027 & $<$2e-16 \\
No$\_$employees$\_$w &- & - & - & 4.004 & 0.039 & $<$2e-16\\
No$\_$subsidiaries$\_$w & 0.216 & 0.031 & 1.83e-12 & 6.242 & 0.058 & $<$2e-16 \\
Op$\_$rev$\_$w & -0.262 & 0.033 & 2.43e-15 &- & - & - \\
PL$\_$beforetax$\_$w & - & - & - & 0.947 & 0.033 & $<$2e-16 \\
Profit$\_$margin$\_$w & 0.106 & 0.022 & 2.08e-06 & 0.261 & 0.017 & $<$2e-16 \\
ROCE$\_$w & - & - & - & 0.439 & 0.010 & $<$2e-16 \\
Sharehold$\_$funds$\_$w & -0.154& 0.023 & 5.90e-11 &- & - & - \\
Solvency$\_$ratio$\_$w & 0.365 & 0.018 & $<$2e-16 & 2.087 & 0.021 & $<$2e-16 \\
Tot$\_$assets$\_$w & - & - & - & 0.750 & 0.024 & $<$2e-16 \\
\hline
\emph{of smooth terms} & \emph{Edf} &  \emph{Est.rank} &  \emph{p-value} & \emph{Edf} &  \emph{Est.rank} &  \emph{p-value}\\
  \hline
Cash$\_$flow$\_$w &- & - & - & 8.808& 9 & $<$ 2e-16\\
Interest$\_$cover$\_$w & 8.488 & 9 & $<$2e-16&- & - & - \\
Net$\_$income$\_$w & 5.592 & 6 & $<$2e-16&- & - & - \\
Op$\_$rev$\_$w &- & - & - & 8.602& 9 & $<$ 2e-16\\
PL$\_$beforetax$\_$w & 8.908 & 9 & $<$2e-16&- & - & - \\
ROCE$\_$w & 8.649 & 9 & $<$2e-16&- & - & - \\
ROE$\_$w &- & - & - & 8.231& 9 & $<$ 2e-16\\
Shareh$\_$liquidity$\_$ratio$\_$w &-& - & - & 7.303& 8 & $<$ 2e-16\\
Sharehold$\_$funds$\_$w &- & - & - & 3.961& 4 & 3.49e-12\\
 \hline
\end{tabular}
\caption{Parametric and smooth component summaries obtained from applying the semiparametric BGEVA model to a sample of Italian and Uk SMEs. The missing values are analysed by Weight of Evidence method. The values of $\tau$ parameter for Italian and Uk models are $-0.41$ and $-0.42$, respectively}
\end{center}
\end{table}

\bibliographystyle{apalike2}
\bibliography{Ref}

\begin{thebibliography}{}

\bibitem[Agresti, 2002]{Agresti:2002}
Agresti, A. (2002).
\newblock {\em Categorical Data Analysis}.
\newblock Wiley, New York.

\bibitem[Altman \& Narayanan, 1997]{Altman:1997}
Altman, E. \& Narayanan, P. (1997).
\newblock An international survey of business failure classification models.
\newblock {\em Financial Markets, Institutions and Instruments}, 6, 1--57.

\bibitem[Altman \& Sabato, 2007]{Altman:2007}
Altman, E. \& Sabato, G. (2007).
\newblock Modeling credit risk for smes: Evidence from the us market.
\newblock {\em ABACUS}, 43(3), 332--357.

\bibitem[Altman et~al., 2010]{Altman:2010}
Altman, E., Sabato, G., \& Wilson, N. (2010).
\newblock The value of non-financial information in small and medium-sized
  enterprise risk management.
\newblock {\em The Journal of Credit Risk}, 6, 1--33.

\bibitem[Becchetti \& Sierra, 2002]{Becchetti:2002}
Becchetti, L. \& Sierra, J. (2002).
\newblock Bankruptcy risk and productive efficiency in manufacturing firms.
\newblock {\em Journal of Banking and Finance}, 27, 2099--2120.

\bibitem[Berg, 2007]{Berg:2007}
Berg, D. (2007).
\newblock Bankruptcy prediction by generalized additive models.
\newblock {\em Applied Stochastic Models in Business and Industry}, 23,
  129--143.

\bibitem[Calabrese et~al., 2013]{Marra:2013}
Calabrese, R., Marra, G., \& Osmetti, S.~A. (2013).
\newblock Bankruptcy prediction of small and medium enterprises using a
  flexible binary generalized extreme value model.
\newblock {\em ArXiv}, 1307.6081, 1--28.

\bibitem[Calabrese \& Osmetti, 2013]{Calabrese:2013}
Calabrese, R. \& Osmetti, S.~A. (2013).
\newblock Modelling sme loan defaults as rare events: the generalized extreme
  value regression model.
\newblock {\em Journal of Applied Statistics}, 40(6), 1172--1188.

\bibitem[Chuang \& Lin, 2009]{Chuang:2009}
Chuang, C.~L. \& Lin, R.~H. (2009).
\newblock Constructing a reassigning credit scoring model.
\newblock {\em Expert Systems with Applications}, 36, 1685--1694.

\bibitem[Ciampi \& Gordini, 2013]{Ciampi:2013}
Ciampi, F. \& Gordini, N. (2013).
\newblock Small enterprise default prediction modeling through artificial
  neural networks: An empirical analysis of italian small enterprises.
\newblock {\em Journal of Small Business Management}, 51(1), 23--45.

\bibitem[Cosh et~al., 1999]{Cosh:1999}
Cosh, A., Hughes, A., \& Wood, E. (1999).
\newblock Innovation in uk smes: causes and consequences for firm failure and
  acquisition.
\newblock In C.~K. Z.~J.~Acs, B.~Carlsson (Ed.), {\em Entrepreneurship, small
  and medium-sized enterprises and the macroeconomy}  (pp.\ 329--366).:
  Cambridge University Press.

\bibitem[Crook, 1996]{Crook:1996}
Crook, J. (1996).
\newblock Credit constraints and us households.
\newblock {\em Applied Financial Economics}, 6(6), 477--485.

\bibitem[Crook et~al., 2007]{Crook:2007}
Crook, J.~N., Edelman, D., \& Thomas, L.~C. (2007).
\newblock Recent developments in consumer credit risk assessment.
\newblock {\em European Journal of Operational Research}, 183, 1447--1465.

\bibitem[Dietsch \& Petey, 2004]{Dietsch:2004}
Dietsch, M. \& Petey, J. (2004).
\newblock Should sme exposure be treated as retail or as corporate exposures? a
  comparative analysis of default probabilities and asset correlation in french
  and german smes.
\newblock {\em Journal of Banking and Finance}, 28, 773--788.

\bibitem[EC, 2012a]{European:2012}
EC (2012a).
\newblock Sba fact sheet 2012 for italy.
\newblock {\em Enterprise and Industry working paper}.

\bibitem[EC, 2012b]{EuropeanUK:2012}
EC (2012b).
\newblock Sba fact sheet 2012 for uk.
\newblock {\em Enterprise and Industry working paper}.

\bibitem[Falk et~al., 2010]{Falk:2010}
Falk, M., Haler, J., \& Reiss, R. (2010).
\newblock {\em Laws of Small Numbers: Extremes and Rare Events}.
\newblock Springer.

\bibitem[Fantazzini \& Figini, 2009]{Fantazzini:2009}
Fantazzini, D. \& Figini, S. (2009).
\newblock Random survival forests models for sme credit risk measurement.
\newblock {\em Methodology and Computing in Applied Probability}, 11, 29--45.

\bibitem[Fantazzini et~al., 2009]{Figini:2009}
Fantazzini, D., Figini, S., Giuli, E.~D., \& Giudici, P. (2009).
\newblock Enhanced credit default models for heterogeneous sme segments.
\newblock {\em Journal of Financial Transformation}, 25(1), 31--39.

\bibitem[Figini \& Giudici, 2012]{Figini:2012}
Figini, S. \& Giudici, P. (2012).
\newblock Statistical merging of rating models.
\newblock {\em Journal of the Operational Research Society}, 62(6), 1067--1074.

\bibitem[Gabrielsson, 2007]{Gabrielsson:2007}
Gabrielsson, J. (2007).
\newblock Boards of directors and entrepreneurial posture in medium-size
  companies. putting the board demography approach to a test.
\newblock {\em International Small Business Journal}, 25, 511--537.

\bibitem[Gestel et~al., 2005]{Gestel:2005}
Gestel, T.~V., Baesens, B., Dijcke, P.~V., Suykens, J. A.~K., Garcia, J., \&
  Alderweireld, T. (2005).
\newblock Linear and non-linear credit scoring by combining logistic regression
  and support vector machines.
\newblock {\em Journal of Credit Risk}, 1(4), 31--60.

\bibitem[Graham, 2012]{Graham:2012}
Graham, J.~W. (2012).
\newblock {\em Missing Data: Analysis and Design}.
\newblock Springer, New York.

\bibitem[Huang et~al., 2006]{Huang:2006}
Huang, J.~J., Tzeng, J.~H., \& Ong, C.~S. (2006).
\newblock Two-stage genetic programming (2sgp) for the credit scoring model.
\newblock {\em Applied Mathematics and Computation}, 174, 1039--1053.

\bibitem[Ihua, 2009]{Ihua:2009}
Ihua, U.~B. (2009).
\newblock Smes key failure-factors: a comparison between the united kingdom and
  nigeria.
\newblock {\em Journal of Social Science}, 18(3), 199--207.

\bibitem[Kiefer, 2010]{Kiefer:2010}
Kiefer, N.~M. (2010).
\newblock Journal of business and economic statistics.
\newblock {\em Journal of Business Finance \& Accounting}, 28(2), 320--328.

\bibitem[King \& Zeng, 2001]{King:2001}
King, G. \& Zeng, L. (2001).
\newblock Logistic regression in rare events data.
\newblock {\em Political Analysis}, 9, 321--354.

\bibitem[Kotz \& Nadarajah, 2000]{Kotz:2000}
Kotz, S. \& Nadarajah, S. (2000).
\newblock {\em Extreme Value Distributions. Theory and Applications}.
\newblock Imperial College Press, London.

\bibitem[Lee \& Chen, 2005]{Lee:2005}
Lee, T.~S. \& Chen, I.~F. (2005).
\newblock A two-stage hybrid credit scoring model using artificial neural
  networks and multivariate adaptive regression splines.
\newblock {\em Expert Systems with Applications}, 28, 743--752.

\bibitem[Lin et~al., 2012]{Lin:2012}
Lin, S.~M., Ansell, J., \& Andreeva, G. (2012).
\newblock Predicting default of a small business using different definitions of
  financial distress.
\newblock {\em Journal of the Operational Research Society}, 63, 539--548.

\bibitem[Lopez, ]{Lopez:2010}
Lopez, R.~F.
\newblock Effects of missing data in credit risk scoring. a comparative
  analysis of methods to achieve robustness in the absence of sufficient data.
\newblock {\em Journal of Operational Research Society}, 61, 486--501.

\bibitem[Lu \& Beamish, 2001]{Lu:2001}
Lu, J.~W. \& Beamish, P.~W. (2001).
\newblock The internationalization and performance of smes.
\newblock {\em Strategic Management Journal}, 22, 565--586.

\bibitem[Lussier \& Halabi, 2010]{Lussier:2010}
Lussier, R. \& Halabi, C. (2010).
\newblock A three-country comparison of the business success versus failure
  prediction model.
\newblock {\em Journal of Small Business Management}, 48(3), 360--377.

\bibitem[Marra et~al., 2013]{bgeva}
Marra, G., Calabrese, R., \& Osmetti, S.~A. (2013).
\newblock {\em bgeva: Binary Generalized Extreme Value Additive Models}.
\newblock R package version 0.2.

\bibitem[Martens et~al., 2011]{Martens:2011}
Martens, D., Vanhoutte, C., Winne, S.~D., Baesens, B., Sels, L., \& Mues, C.
  (2011).
\newblock Identifying financially successful start-up profiles with data
  mining.
\newblock {\em Expert Systems with Applications}, 38, 5794--5800.

\bibitem[Michala et~al., 2013]{Michala:2013}
Michala, D., Grammatikosa, T., \& Filipea, S.~F. (2013).
\newblock Forecasting distress in european sme portfolios.
\newblock {\em LSF Research Working Paper Series}, (13-02).

\bibitem[Ong et~al., 2005]{Ong:2005}
Ong, C.~S., Huanga, J.~J., \& Tzeng, G.~H. (2005).
\newblock Building credit scoring models using genetic programming. expert
  systems with applications.
\newblock {\em Expert Systems with Applications}, 29, 41--47.

\bibitem[Orton et~al., 2011]{Orton:2011}
Orton, P., Ansell, J., \& Andreeva, G. (2011).
\newblock Recent developments in commercial scoring.
\newblock {\em Credit Scoring \& Credit Control XII Conference}, Edinburgh, UK.

\bibitem[Pederzoli et~al., 2013]{Pederzoli:2013}
Pederzoli, C., Thoma, G., \& Torriccelli, C. (2013).
\newblock Modelling credit risk for innovative smes: the role of innovation
  measures.
\newblock {\em Journal of Financial Services Research}, 44, 111--129.

\bibitem[Rubin, 1977]{Rubin:1977}
Rubin, D.~B. (1977).
\newblock Formalizing subjective notions about the effect of nonrespondents in
  sample surveys.
\newblock {\em Journal of the American Statistical Association}, 72, 538–543.

\bibitem[Ruppert et~al., 2003]{Ruppert:2003}
Ruppert, D., Wand, M.~P., \& Carroll, R.~J. (2003).
\newblock {\em Semiparametric Regression}.
\newblock Cambridge University Press, London.

\bibitem[Shumway, 2001]{Shumway:2001}
Shumway, T. (2001).
\newblock Forecasting bankruptcy more accurately: A simple hazard model.
\newblock {\em Journal of Business}, 74(1), 101—124.

\bibitem[Sohn \& Kim, 2013]{Sohn:2013}
Sohn, S.~Y. \& Kim, Y.~S. (2013).
\newblock Behavioral credit scoring model for technology-based firms that
  considers uncertain financial ratios obtained from relationship banking.
\newblock {\em Small Business Economics}, 41(4), 931--943.

\bibitem[Thomas et~al., 2002]{Thomas:2002}
Thomas, L., Edelman, D., \& Crook, J.~C. (2002).
\newblock {\em Credit Scoring and Its Applications}.
\newblock Society for Industrial and Applied Mathematics, Philadelphia.

\bibitem[Twala, 2009]{Twala:2009}
Twala, B. (2009).
\newblock Combining classifiers for credit risk prediction.
\newblock {\em Journal of Systems Science and Systems Engineering}, 18(3),
  292--311.

\bibitem[Vallini et~al., 2009]{Vallini:2009}
Vallini, C.~F., Ciampi, F., Gordini, N., \& Benvenuti, M. (2009).
\newblock Are credit scoring model able to predict small enterprise default?
  statistical evidence from italian firms.
\newblock {\em International Journal of Business \& Economics}, 8(1), 3--18.

\bibitem[Wang \& Dey, 2010]{Wang2010}
Wang, X. \& Dey, D.~K. (2010).
\newblock Generalized extreme value regression for binary response data: An
  application to b2b electronic payments system adoption.
\newblock {\em The Annals of Applied Statistics}, 4(4), 2000--2023.

\bibitem[Wood, 2006]{Wood:2006}
Wood, S.~N. (2006).
\newblock {\em Generalized Additive Models: An Introduction with R}.
\newblock Chapman $\&$ Hall, Boca Raton.

\bibitem[Zavgren, 1998]{Zavgren:1998}
Zavgren, C. (1998).
\newblock The prediction of corporate failure: the state of the art.
\newblock {\em Journal of Accounting Literature}, 2, 1--37.

\end{thebibliography}
\end{document}